\documentstyle[12pt]{article}
\begin{document}
\begin{center}
{\bf{THE TWO AND THREE POINT ONE LOOP FUNCTIONS}}\footnote{The
research described in this publication was possible in part by Grant
No. RVH 280 from The Joint Program of Georgia Goverment and International
Science Foundation}
\  \par
\   \par
{\bf {G.G.~Devidze}}\footnote{Present address: Laboratory of theoretical
Physics, Joint Institute for Nuclear Research, 141980 Dubna, Moscow Region,
Russia} , {\bf {G.R.~Jibuti}}\\
 
\    \par
High Energy Physics Institute\\
Tbilisi State University\\
University St. 9, 380086 Tbilisi, Rep. of Georgia\\
E-mail: devidze@hepi.edu.ge\\
E-mail: jibuti@hepi.edu.ge
\end{center}
\ \par
\ \par

In the present paper the two and three point functions,
which occur at the study of the various physical processes are considered.
The investigation is done in the framework of the perturbation theory
at the one loop level.  The general analytical and asymptotic expressions for
these functions are obtained.
\newpage
\begin{center}
\section {INTRODUCTION.}
\end{center}
\  \par
 
  With the increasing role which place the perturbative aprouche within
 the field theory, the evolution of the effects beyond tree approximation
 in the elementary particle physics becoming of great interests. Such
 effects play an important role e.g. in the investigation of rare decays of
 hadrons, leptons,quarks, meson oscilation ets. Within the framework of standard
 theory of electro weak interactions.
 
 During the study of physical processes beyond the tree approximation within
 the perturbative aprouche imarge so called loop integrals.The study of loop
 integrals and the investigation of the behavior of this integrals for the
 various values of cinematic parameters is over great interest.Such investigation
 have been cared out in references [1-5].However for the practical point of weu
 the utilization of the results obtained in these paper encounters the
 difficulties. In the present work the two and tree point functions which
 appear during the study of the physical processes at the one loop level are
 investigated. The general and asymptotic expressions for these functions for
 the different values of cinematic parameters are obtained.
 
\  \par
\begin{center}
\section {INTEGRAL EXPRESSIONS OF THE TWO AND TREE POINT FUNCTIONS.}
\end{center}
\  \par
 
 During the study physical processes at one loop level there exist integrals
 of the following types:
$$
 \{I_{0};I_{\alpha}\}(p^{2},m_{1}^{2},m_{2}^{2})=
 \int \frac{d^{4}q}{(2\pi)^{4}}
 \frac{(1,q_{\alpha})}{[(q-p)^{2}-m^{2}_{1}+i\epsilon]
 (q^{2}-m^{2}_{2}+i\epsilon)}~~~~~,  \eqno{(1)}
$$
$$
\{R_{0},R_{\alpha},R_{\alpha\beta)}\}(p^{2}_{1},p^{2}_{2},(p_{1}-p_{2})^{2},
 m^{2}_{1},m^{2}_{2},m^{2}_{3})=
$$
$$
 \int \frac{d^{4}q}{(2\pi)^{4}}\frac{(1,q_{\alpha},q_{\alpha}
 q_{\beta})}{[(p_{1}-q)^{2}-m_{1}^{2}+i\epsilon]
 [(p_{2}-q)^{2}-m_{2}^{2}+i\epsilon]
 (q^{2}-m_{3}^{2}+i\epsilon)}~~~~~.\eqno{(2)}
$$
 
 The Lorentzdecomposition of these integrals
 ( $I_{\alpha}$,~$R_{\alpha}$ and
 $R_{\alpha\beta}$) have folowing forms:
$$
  I_{\alpha}(p^{2},m^{2}_{1},m^{2}_{2})=p_{\alpha}I_{1}(p^{2},
 m^{2}_{1},m^{2}_{2})~~,
$$
$$
  R_{\alpha}(p^{2}_{1},p^{2}_{2},(p_{1}-p_{2})^{2},
 m^{2}_{1},m^{2}_{2},m^{2}_{3})=
 p_{1\alpha}R_{10}(p^{2}_{1},p^{2}_{2},(p_{1}-p_{2})^{2}
 m^{2}_{1},m^{2}_{2},m^{2}_{3})+
 $$
 $$
 p_{2\alpha}R_{01}(p^{2}_{1},p^{2}_{2},(p_{1}-p_{2})^{2}
 m^{2}_{1},m^{2}_{2},m^{2}_{3})~~,
$$
$$
  R_{\alpha\beta}(p^{2}_{1},p^{2}_{2},(p_{1}-p_{2})^{2},
 m^{2}_{1},m^{2}_{2},m^{2}_{3})=
 g_{\alpha\beta}R_{00}(p^{2}_{1},p^{2}_{2},(p_{1}-p_{2})^{2},
 m^{2}_{1},m^{2}_{2},m^{2}_{3})+
$$
$$
  p_{1\alpha}p_{1\beta}R_{20}(p^{2}_{1},p^{2}_{2},(p_{1}-p_{2})^{2},
 m^{2}_{1},m^{2}_{2},m^{2}_{3})+
 p_{2\alpha}p_{2\beta}R_{02}(p^{2}_{1},p^{2}_{2},(p_{1}-p_{2})^{2},
 m^{2}_{1},m^{2}_{2},m^{2}_{3})+
$$
$$
  (p_{1\alpha}p_{2\beta}+p_{2\alpha}p_{1\beta})
 R_{11}(p^{2}_{1},p^{2}_{2},(p_{1}-p_{2})^{2},
 m^{2}_{1},m^{2}_{2},m^{2}_{3})~~. \eqno{(3)}
$$
 Some of these integralas contaain divergences. In the dimenshinal regularization
 scheem it is possible to single out the divergent part in the following form:
$$
 I_{0}(p^{2},m_{1}^{2},m_{2}^{2})=\frac{1}{\epsilon^{'}}+
 \tilde I_{0}(p^{2},m_{1}^{2},m_{2}^{2})~~~~~,
$$
$$
 I_{\alpha}(p^{2},m^{2}_{1},m^{2}_{2})=p_{\alpha}I_{1}(p^{2},
 m^{2}_{1},m^{2}_{2})=
 p_{\alpha}\{\frac{1}{2\epsilon^{'}}+
 \tilde I_{1}(p^{2},m^{2}_{1},m^{2}_{2})\}~~~~~,
$$
$$
 R_{00}(p^{2}_{1},p^{2}_{2},(p_{1}-p_{2})^{2},
 m^{2}_{1},m^{2}_{2},m^{2}_{3})=
 \frac{1}{4\epsilon^{'}}+
 \tilde R_{00}(p^{2}_{1},p^{2}_{2},(p_{1}-p_{2})^{2},m_{1}^{2},m_{2}^{2},
 m_{3}^{2})~, \eqno{(4)}
$$
 where
 $1/\epsilon^{'}=\frac {i}{(4\pi)^{2}}[1/\epsilon-\gamma+\ell n4\pi]$,
 $\gamma$ is eiler maskeroni constant,
 ($\gamma=.5772157...$),
 $2\epsilon=4-n$, n is dimenshin of space-time.
 In the framework of Feinman parametrization the finite parts of these
 integrals have respectively the form:
$$
\tilde
 I_{0}(p^{2},m^{2}_{1},m^{2}_{2})= -\frac{i}{(4\pi)^{2}}\int \limits_{0}^{1} dx
 \ell n\frac{m_{1}^{2}(1-x)+m_{2}^{2}x-p^{2}x(1-x)+i\epsilon}{\mu^{2}}~~,
 \eqno{(5)}
$$
$$
  \tilde I_{1}(p^{2},m^{2}_{1},m^{2}_{2})=
 -\frac{i}{(4\pi)^{2}}\int \limits_{0}^{1} dx
 (1-x)\ell n\frac{
 m_{1}^{2}(1-x)+m_{2}^{2}x-p^{2}x(1-x)+i\epsilon}{\mu^{2}}~~, \eqno{(6)}
$$
$$
 \tilde
 R_{00}(p^{2}_{1},p^{2}_{2},(p_{1}-p_{2})^{2},m^{2}_{1},m^{2}_{2},m^{2}_{3})=
$$
$$
 -\frac{i}{(4\pi)^{2}}\int \limits_{0}^{1}dx
 \int
 \limits_{0}^{1}dy\frac{x}{2} \ell n\frac{
 L(p^{2}_{1},p^{2}_{2},(p_{1}-p_{2})^{2},m^{2}_{1},m^{2}_{2},m^{2}_{3},x,y)}
 {\mu^{2}}~~,\eqno{(7)}
$$
$$
  R_{ij}(p^{2}_{1},p^{2}_{2},(p_{1}-p_{2})^{2},m^{2}_{1},m^{2}_{2},m^{2}_{3})=
 -\frac{i}{(4\pi)^{2}}\int \limits_{0}^{1}dx
 \int
 \limits_{0}^{1}dy\frac{x^{3}y^{j}(1-y)^{i}}
 {L(p^{2}_{1},p^{2}_{2},(p_{1}-p_{2})^{2},m^{2}_{1},m^{2}_{2},m^{2}_{3},x,y)}~~,
$$
$$
  (i+j=2;~~i,j=0,1,2) \eqno{(8)}
$$
$$
  R_{ij}(p^{2}_{1},p^{2}_{2},(p_{1}-p_{2})^{2},m^{2}_{1},m^{2}_{2},m^{2}_{3})=
 -\frac{i}{(4\pi)^{2}}\int \limits_{0}^{1}dx
 \int
 \limits_{0}^{1}dy\frac{x^{2}y^{1-i}(1-y)^{1-j}}
 {L(p^{2}_{1},p^{2}_{2},(p_{1}-p_{2})^{2},m^{2}_{1},m^{2}_{2},m^{2}_{3},x,y)}~~,
$$
$$
  (i+j=1;~~i,j=0,1) \eqno{(9)}
$$
$$
  R_{0}(p^{2}_{1},p^{2}_{2},(p_{1}-p_{2})^{2},m^{2}_{1},m^{2}_{2},m^{2}_{3})=
$$
$$
 -\frac{i}{(4\pi)^{2}}\int \limits_{0}^{1}dx
 \int
 \limits_{0}^{1}dy\frac{x}
 {L(p^{2}_{1},p^{2}_{2},(p_{1}-p_{2})^{2},m^{2}_{1},m^{2}_{2},m^{2}_{3},x,y)}
 ~~,\eqno{(10)}
$$
 where $\mu^{2}$ is arbitrary mass scale parameter and
 we have introduce the following notations:
$$
 L(p^{2}_{1},p^{2}_{2},(p_{1}-p_{2})^{2},m^{2}_{1},m^{2}_{2},m^{2}_{3},x,y)=
 Ax^{2}+Bx+m^{2}_{2}~~,
$$
$$
     A=[p_{2}+y(p_{1}-p_{2})]^{2},
$$
$$
     B=(m_{1}^{2}-i\epsilon )y+(m_{2}^{2}-i\epsilon )(1-y)-
     m_{3}^{2}+i\epsilon -p_{1}^{2}y - p_{2}^{2}(1-y).
$$
\begin{center}
\section {TWO POINT FUNCTIONS.}
\end{center}
\  \par
 
 Integrals
 $\tilde I_{0}(p^{2},m^{2}_{1},m^{2}_{2})$ and
  $\tilde I_{1}(p^{2},m^{2}_{1},m^{2}_{2})$,
 after integration by part can take in the followingb tipe:
$$
\tilde I_{0}(p^{2},m^{2}_{1},m^{2}_{2})=
 -\frac {i}{(4\pi)^{2}} \{\ell n\frac {m^{2}_{2}}{\mu^{2}}-
 \int_{0}^{1}dxx\frac {m_{2}^{2}-m^{2}_{1}-p^{2}+2p^{2}x}
 {m_{1}^{2}(1-x)+m_{2}^{2}x-p^{2}x(1-x)-i\epsilon}  \}~,\eqno{(11)}
$$
$$
 \tilde I_{1}(p^{2},m^{2}_{1},m^{2}_{2})=
 -\frac {i}{(4\pi)^{2}} \{ \frac {1}{2} \ell n\frac {m^{2}_{2}}{\mu^{2}}-
 \int_{0}^{1}dxx(1-\frac {x}{2})\frac {m_{2}^{2}-m^{2}_{1}-p^{2}+2p^{2}x}
 {m_{1}^{2}(1-x)+m_{2}^{2}x-p^{2}x(1-x)-i\epsilon}\}~.\eqno{(12)}
$$
 
 The explisit expressions
 $\tilde I_{0}$ and $\tilde I_{1}$
 for the diferent cinematic of the physical processes after some calculations
 become the following form:
$$
 \tilde I_{0}(p^{2},m^{2}_{1},m^{2}_{2})=
 -\frac {i}{(4\pi)^{2}} \{\ell n\frac {m^{2}_{2}}{\mu^{2}}+
 \frac {m_{2}^{2}-m_{1}^{2}-p^{2}}{2p^{2}}\ell n\frac {m^{2}_{2}}{m_{1}^{2}}-
 2+
$$
$$
 \frac {\sqrt {-\lambda (m_{1}^{2},m^{2}_{2},p^{2})}}{p^{2}}arctg
 \frac {\sqrt {-\lambda (m_{1}^{2},m^{2}_{2},p^{2})}}{m_{2}^{2}+m_{1}^{2}-p^{2}}
 \}~,
$$
$$
 \tilde I_{1}(p^{2},m^{2}_{1},m^{2}_{2})=
 -\frac {i}{(4\pi)^{2}} \{ \frac {1}{2}\ell n\frac {m^{2}_{2}}{\mu^{2}}-
 \frac {3}{2}-
 \frac {m_{2}^{2}-m_{1}^{2}-p^{2}}{2p^{2}}+
$$
$$
 \frac {(m_{2}^{2}-m_{1}^{2}-p^{2})^{2}+2p^{2}(m_{2}^{2}-2m_{1}^{2}-p^{2})
 }{4p^{4}}\ell n\frac {m^{2}_{2}}{m_{1}^{2}}+
$$
$$
 \frac {m_{2}^{2}-m_{1}^{2}-3p^{2}
 \sqrt {-\lambda (m_{1}^{2},m^{2}_{2},p^{2})}}
 {4p^{4}}arctg
 \frac {\sqrt {-\lambda (m_{1}^{2},m^{2}_{2},p^{2})}}
 {m_{2}^{2}+m_{1}^{2}-p^{2}}\}~, \eqno{(13)}
$$
 if
 $(m_{1}-m_{2})^{2}<p^{2}<(m_{1}+m_{2})^{2}$;
 here
 $\lambda (x,y,z)=x^{2}+y^{2}+z^{2}-2xy-2zy-2xz$,
$$
 \tilde I_{0}(p^{2},m^{2}_{1},m^{2}_{2})=
 -\frac {i}{(4\pi)^{2}} \{\ell n\frac {m^{2}_{2}}{\mu^{2}}-
 i\pi S_{0}+
 \frac {m_{2}^{2}-m_{1}^{2}-p^{2}}{2p^{2}}\ell n\frac {m^{2}_{2}}{m_{1}^{2}}-
 2-
$$
$$
 \frac {\sqrt {\lambda (m_{1}^{2},m^{2}_{2},p^{2})}}{2p^{2}}\ell n
 \frac {(m_{1}^{2}+m_{2}^{2}-p^{2}+
 \sqrt {\lambda (m_{1}^{2},m^{2}_{2},p^{2})})^{2}}{4m_{1}^{2}m_{2}^{2}}
 \}~~,
$$
$$
 \tilde I_{1}(p^{2},m^{2}_{1},m^{2}_{2})=
 -\frac {i}{(4\pi)^{2}} \{ \frac {1}{2}\ell n\frac {m^{2}_{1}}{\mu^{2}}-
 i\pi S_{1}+
 \frac {m_{1}^{2}-m_{2}^{2}-p^{2}}{2p^{2}}-
 \frac {1}{2}-
 \frac {(m_{1}^{2}-m_{2}^{2}-p^{2})^{2}-2p^{2}m_{2}^{2}}
 {4p^{4}}\ell n\frac {m^{2}_{1}}{m_{2}^{2}}+
$$
$$
 \frac {m_{1}^{2}-m_{2}^{2}-p^{2}
 \sqrt {\lambda (m_{1}^{2},m^{2}_{2},p^{2})}}
 {4p^{4}}\ell n
 \frac {(m_{1}^{2}+m_{2}^{2}-p^{2}+
 \sqrt {\lambda (m_{1}^{2},m^{2}_{2},p^{2})})^{2}}
 {4m_{1}^{2}m_{2}^{2}}\}~~, \eqno{(14)}
$$
 if
 $p^{2}\leq (m_{1}-m_{2})^{2}\bigcup~~
 p^{2}\geq (m_{1}+m_{2})^{2}$.
 
 In the formula (14) $S_{0}$ and $S_{1}$ are the contribution due the pole of
 the integrals (11),(12):
$$
 S_{0}=
 \left( \begin{array}{ll}
 \sqrt {\lambda (m_{1}^{2},m^{2}_{2},p^{2})}/p^{2}; &
 ~~ x_{1}\in [0,1]~~ ,~~ x_{2}\in [0,1],  \\
 (m_{1}^{2}-m_{2}^{2}+p^{2}+
 \sqrt {\lambda (m_{1}^{2},m^{2}_{2},p^{2})})/2p^{2}; &
 ~~ x_{1}\in [0,1]~~ ,~~ x_{2}\not\in [0,1],  \\
 (m_{2}^{2}-m_{1}^{2}-p^{2}+
 \sqrt {\lambda (m_{1}^{2},m^{2}_{2},p^{2})})/2p^{2}; &
 ~~ x_{1}\not\in [0,1]~~ ,~~ x_{2}\in [0,1],  \\
 0; &
 ~~ x_{1}\not\in [0,1]~~ ,~~ x_{2}\not\in [0,1],
 \end{array} \right. \eqno{(15)}
$$
$$
 S_{1}=
 \left( \begin{array}{ll}
 (m_{2}^{2}-m_{1}^{2}+p^{2})
 \sqrt {\lambda (m_{1}^{2},m^{2}_{2},p^{2})}/2p^{4}; &
 ~~ x_{1}\in [0,1]~~ ,~~ x_{2}\in [0,1],  \\
 \frac {1}{4p^{4}}[(m_{1}^{2}-m_{2}^{2})^{2}+p^{4}-
 2m_{1}^{2}p^{2}-(m_{1}^{2}-\\
 m_{2}^{2}-p^{2})
 \sqrt {\lambda (m_{1}^{2},m^{2}_{2},p^{2})}]; &
 ~~ x_{1}\in [0,1]~~ ,~~ x_{2}\not\in [0,1],  \\
 \frac {1}{4p^{4}}[-(m_{1}^{2}-m_{2}^{2})^{2}-p^{4}+
 2m_{1}^{2}p^{2}-(m_{1}^{2}- \\
 m_{2}^{2}-p^{2})
 \sqrt {\lambda (m_{1}^{2},m^{2}_{2},p^{2})}]; &
 ~~ x_{1}\not\in [0,1]~~ ,~~ x_{2}\in [0,1],  \\
 0; &
 ~~ x_{1}\not\in [0,1]~~ ,~~ x_{2}\not\in [0,1],
 \end{array} \right. \eqno{(16)}
$$
 where $x_{1}$ and $x_{2}$ are the roots of quadratic polynomial
 $p^{2}x^{2}+(m_{2}^{2}-m_{1}^{2}-p^{2})x+m_{1}^{2}$.
 It shoold be mention.that in integrals (11) and (12) ,when
 the cinematic $p^{2}\leq (m_{1}-m_{2})^{2}\bigcup~~
 p^{2}\geq (m_{1}+m_{2})^{2}$ , it appears poles in defend of
 following parameters $m_{1}^{2}$,$m_{2}^{2}$, $p^{2}$.
 During the investigation of physical processes, when one or more
 parameters are significantly large in compare with others, it is more
 conveniently to use approxomately expressions of integrals
 $\tilde I_{0}$ and $\tilde I_{1}$.
 We will show below approximately expressions of these integrals for
 frequantly arising values of kinematic parameters
 ($p^{2}<m_{1}^{2}$ and/or $p^{2}<m_{2}^{2}$):
$$
 \tilde
 I_{0}(p^{2},m^{2}_{1},m^{2}_{2})= -\frac {i}{(4\pi)^{2}} \{\ell n\frac
 {m^{2}_{1}}{\mu^{2}}+ a_{1}(x)+\frac {p^{2}}{m_{1}^{2}}a_{2}(x)+\frac
 {p^{4}}{m_{1}^{4}}a_{3}(x)\}~,
$$
$$
 \tilde
 I_{1}(p^{2},m^{2}_{1},m^{2}_{2})= -\frac {i}{(4\pi)^{2}} \{\frac {1}{2}\ell
 n\frac {m^{2}_{1}}{\mu^{2}}+ b_{1}(x)+\frac {p^{2}}{m_{1}^{2}}b_{2}(x)+\frac
 {p^{4}}{m_{1}^{4}}b_{3}(x)\}~, \eqno{(17)}
$$
 Where we apply following notation:
$$
 k_{1}(x)=\frac {x-1-x\ell n x}{1-x},~~~ k_{2}(x)=\frac
 {x^{2}-1-2x\ell n x}{2(1-x)^{3}},~~ k_{3}(x)=\frac
 {x^{3}+9x^{2}-9x-1-6x(1+x)\ell n x}{6(1-x)^{5}},
$$
$$
 l_{1}(x)=\frac
 {-3x^{2}+4x-1+2x^{2}\ell n x}{4(1-x)^{2}},~~ l_{2}(x)=\frac
 {-2x^{3}-3x^{2}+6x-1+6x^{2}\ell n x}{6(1-x)^{4}},
$$
$$
 l_{3}(x)=\frac
 {-3x^{4}-44x^{3}+36x^{2}+12x-1+12 (3x^{2}+2x^{3})\ell n x}{24(1-x)^{6}},~~
 x=\frac {m_{2}^{2}}{m_{1}^{2}}. \eqno{(17')}
$$
 \par
 In the framework of large momentum
 $(p^{2}-m_{1}^{2}>m_{2}^{2})$, the approximative expressions of
 $\tilde I_{0}$ and $\tilde I_{1}$ are obtained the folowing form
 consequently:
$$
 \tilde I_{0}(p^{2},m^{2}_{1},m^{2}_{2})=
 -\frac {i}{(4\pi)^{2}} \{\ell n\frac {m^{2}_{2}}{\mu^{2}}-
 i\pi[1-\frac {m_{1}^{2}}{p^{2}}-\frac {m_{2}^{2}(p^{2}+m_{1}^{2})}
 {p^{2}(p^{2}-m_{1}^{2})}]-2+
$$
$$
 \frac {m_{2}^{2}-p^{2}+m_{1}^{2}}
 {p^{2}-m_{1}^{2}}\ell n\frac {m_{2}^{2}}{m_{1}^{2}}+
 [1-\frac {m_{1}^{2}}{p^{2}}-\frac {m_{2}^{2}(p^{2}+m_{1}^{2})}
 {p^{2}(p^{2}-m_{1}^{2})}]\ell n(\frac {p^{2}}{m_{1}^{2}}-1)\}
$$
$$
 \tilde I_{1}(p^{2},m^{2}_{1},m^{2}_{2})=
 -\frac {i}{(4\pi)^{2}} \{-\frac {1}{2}\ell n\frac {m^{2}_{1}}{\mu^{2}}+
 \frac {m_{1}^{2}}{p^{2}}\ell n\frac {m^{2}_{1}}{\mu^{2}}+
 \frac {i\pi}{2}[1-2\frac {m_{1}^{2}}{p^{2}}(1+\frac {m_{2}^{2}}
 {p^{2}-m_{1}^{2}})+
$$
$$
 \frac {m_{1}^{4}}{p^{4}}(1+\frac {2m_{2}^{2}}
 {p^{2}-m_{1}^{2}})]+1-\frac {3m_{1}^{2}}{2p^{2}}+
 \frac {3m_{2}^{2}}{2p^{2}}-\frac {m_{2}^{2}}{p^{2}}\ell n\frac{m_{2}^{2}}
 {m_{1}^{2}}-
$$
$$
 \frac {1}{2}[1-2\frac {m_{1}^{2}}{p^{2}}(1+\frac {m_{2}^{2}}
 {p^{2}-m_{1}^{2}})+\frac {m_{1}^{4}}{p^{4}}(1+\frac {2m_{2}^{2}}
 {p^{2}-m_{1}^{2}})]\ell n(\frac {p^{2}}{m_{1}^{2}}-1)\} \eqno{(18)}
$$
 
 It should be pointed out that at the zero value of the cinematic
 parameters the formulae (14), (17) and (18) should be understood
 as limited expressions.
 
 Below we give usefull relations for the two point functions
$$
 I_{0}(p^{2},m^{2}_{1},m^{2}_{2})=
 I_{0}(p^{2},m^{2}_{2},m^{2}_{1}),
$$
$$
 I_{1}(p^{2},m^{2}_{1},m^{2}_{2})=
 I_{0}(p^{2},m^{2}_{2},m^{2}_{1})-
 I_{1}(p^{2},m^{2}_{2},m^{2}_{1}).
$$
 We have the same relations for the functions
 $\tilde I_{0}$ and $\tilde I_{1}$.
\begin{center}
\section {TREE POINT FUNCTIONS.}
\end{center}
\  \par
 
 Let us go to the more detail investigation of the tree point functions
 $R_{0}$,$R_{\alpha}$ and $R_{\alpha\beta}$.
 Multiplying the second expression in the formulae (3) on
 $p_{1\alpha}$ and $p_{2\alpha}$ we obtain the system of equations for the
 functions $R_{10}$ and $R_{01}$:
$$
 \left( \begin{array}{ccc}
 2p_{1}^{2}R_{10}+2(p_{1}p_{2})R_{01}&=&
 I_{0}((p_{1}-p_{2})^{2},m_{1}^{2},m_{2}^{2})-
 I_{0}(p_{2}^{2},m_{2}^{2},m_{3}^{2})+\\
 & &(m_{3}^{2}-m_{1}^{2}+p_{1}^{2})R_{0},\\
 2(p_{1}p_{2})R_{10}-2p_{2}^{2}R_{01}&=&
 I_{0}((p_{1}-p_{2})^{2},m_{1}^{2},m_{2}^{2})-
 I_{0}(p_{1}^{2},m_{1}^{2},m_{3}^{2})+\\
 & &(m_{3}^{2}-m_{2}^{2}+p_{2}^{2})R_{0},
 \end{array}     \right. \eqno{(19)}
$$
 
 After solving the system (19) we get following expressions for
 the tree point functions $R_{10}$ and $R_{01}$:
$$
 R_{10}=\frac {1}{2[p_{1}^{2}p_{2}^{2}-(p_{1}p_{2})^{2}]}
 \{p_{2}^{2}[I_{0}(p_{2}^{2},m_{2}^{2},m_{3}^{2})-
 I_{0}((p_{1}-p_{2})^{2},m_{1}^{2},m_{2}^{2})]+
$$
$$
 (p_{1}p_{2})[I_{0}((p_{1}-p_{2})^{2},m_{1}^{2},m_{2}^{2})-
 I_{0}(p_{1}^{2},m_{1}^{2},m_{3}^{2})]+
$$
$$
 [p_{2}^{2}(m_{1}^{2}-m_{3}^{2}-p_{1}^{2})-
 (p_{1}p_{2})(m_{2}^{2}-m_{3}^{2}-p_{2}^{2})]R_{0}\},
$$
$$
 R_{01}=\frac {1}{2[p_{1}^{2}p_{2}^{2}-(p_{1}p_{2})^{2}]}
 \{p_{1}^{2}[I_{0}(p_{1}^{2},m_{1}^{2},m_{3}^{2})-
 I_{0}((p_{1}-p_{2})^{2},m_{1}^{2},m_{2}^{2})]+
$$
$$
 (p_{1}p_{2})[I_{0}((p_{1}-p_{2})^{2},m_{1}^{2},m_{2}^{2})-
 I_{0}(p_{2}^{2},m_{2}^{2},m_{3}^{2})]+
$$
$$
 [p_{1}^{2}(m_{2}^{2}-m_{3}^{2}-p_{2}^{2})-
 (p_{1}p_{2})(m_{1}^{2}-m_{3}^{2}-p_{1}^{2})]R_{0}\}. \eqno{(20)}
$$
 
 In order to obtain explicit expressions for the functions
 $R_{00}$, $R_{20}$, $R_{02}$, $R_{11}$ let us multiply the third
 term of formula (3) on $p_{1\alpha}$ and $p_{2\alpha}$.
 In results we get the system of linear equations, whose determinant
 is equal zero. We can add one independent equation obtained via the
 multiplying of the Lorentz decomposition of $R_{\alpha\beta}$
 from formula (3) on $g_{\alpha\beta}$.
 The new system can be explicitly solved with respect to the functions
 $R_{00}$, $R_{20}$, $R_{02}$, $R_{11}$ and has the form:
$$
 \left( \begin{array}{ccc}
 & 4R_{00}+p_{1}^{2}R_{20}+2(p_{1}p_{2})R_{11}+p_{2}^{2}R_{02}=
 m_{3}^{2}R_{0}+I_{0}((p_{1}-p_{2})^{2},m_{1}^{2},m_{2}^{2})\equiv a_{1}\\
 & R_{00}+p_{1}^{2}R_{20}+(p_{1}p_{2})R_{11}=-\frac {1}{2}\{
 (m_{1}^{2}-m_{3}^{2}-p^{2}_{1})R_{10}-
 I_{1}((p_{1}-p_{2})^{2},m_{1}^{2},m_{2}^{2})\}\equiv a_{2} \\
 & p_{1}^{2}R_{11}+(p_{1}p_{2})R_{02}=-\frac {1}{2}\{
 (m_{1}^{2}-m_{3}^{2}-p^{2}_{1})R_{01}+
 I_{1}((p_{1}-p_{2})^{2},m_{1}^{2},m_{2}^{2})- \\
 & I_{0}((p_{1}-p_{2})^{2},m_{1}^{2},m_{2}^{2})+
 I_{1}(p_{2}^{2},m_{2}^{2},m_{3}^{2})\}\equiv a_{3} \\
 & (p_{1}p_{2})R_{20}+p_{2}^{2}R_{11}=-\frac{1}{2}\{
 (m_{2}^{2}-m_{3}^{2}-p^{2}_{2})R_{10}-
 I_{1}((p_{1}-p_{2})^{2},m_{1}^{2},m_{2}^{2})+ \\
 & I_{1}(p_{1}^{2},m_{1}^{2},m_{3}^{2})\} \equiv a_{4} .
\end{array} \right. \eqno{(21)}
$$
\par\
\par\
 
 The solution of the given above system of equation has the following form:
$$
 R_{00}=\frac {1}{2}\{a_{1}-2a_{2}-\frac {p_{2}^{2}}{(p_{1}p_{2})}a_{3}+
 \frac {p_{1}^{2}}{(p_{1}p_{2})}a_{4}\},
$$
$$
R_{20}=\frac {a_{1}p_{2}^{2}(p_{1}p_{2})-4a_{2}p_{2}^{2}(p_{1}p_{2})-
 a_{3}p_{2}^{4}+a_{4}(2(p_{1}p_{2})^{2}+p_{1}^{2}p_{2}^{2})}
 {2(p_{1}p_{2})[(p_{1}p_{2})^{2}-p_{1}^{2}p_{2}^{2}]}~~,
$$
$$
 R_{11}=\frac {(p_{1}p_{2})(-a_{1}+4a_{2})+p_{2}^{2}a_{3}-3p_{1}^{2}a_{4}}
 {2[(p_{1}p_{2})^{2}-p_{1}^{2}p_{2}^{2}]}~~,
$$
$$
 R_{02}=\frac {(a_{1}-4a_{2})p_{1}^{2}(p_{1}p_{2})+a_{3}[2(p_{1}p_{2})^{2}-
 3p_{1}^{2}p_{2}^{2}]+3a_{4}p_{1}^{4}}
 {2(p_{1}p_{2})[(p_{1}p_{2})^{2}-p_{1}^{2}p_{2}^{2}]}. \eqno{(22)}
$$
 
>From the expressions (20), (22), given above one observes that all three
 point functions can be expressed via the two point functions
 $I_{0}$,$I_{1}$ (whose explicit expressions are given in section (3)
 (see formulae (13) and (14)) and the three point function $R_{0}$.
 Bellow we shall obtained the explicit expressions for the three point
 function$R_{0}$.
\begin{center}
\section {CALCULATION OF THE THREE POINT FUNCTION $R_{0}$}.
\end{center}
 
 Let us consider the three point function $R_{0}$:
$$
 R_{0}(p^{2}_{1},p^{2}_{2},(p_{1}-p_{2})^{2},
 m^{2}_{1},m^{2}_{2},m^{2}_{3})=
$$
$$
 \int \frac{d^{4}q}{(2\pi)^{4}}\frac{(1,q_{\alpha},q_{\alpha}
 q_{\beta})}{[(p_{1}-q)^{2}-m_{1}^{2}+i\epsilon]
 [(p_{2}-q)^{2}-m_{2}^{2}+i\epsilon]
 (q^{2}-m_{3}^{2}+i\epsilon)}~~. \eqno{(23)}
$$
 After using the Feinman parametrization the three point function
 (23) is written in the following form:
$$
 R_{0}=\frac {i}{16\pi^{2}}\int \limits_{0}^{1}dx\int \limits_{0}^{x}dy
 \frac {1}{\{ax^{2}+by^{2}+cxy+dx+ey+f\}}~~, \eqno{(24)}
$$
 where
 $a=-p_{2}^{2}$, $b=-(p_{1}-p_{2})^{2}$, $c=2p_{2}^{2}-2(p_{1}p_{2})$,
 $d=p_{2}^{2}-m_{2}^{2}+m_{3}^{2}$, $e=p_{1}^{2}-p_{2}^{2}-m_{1}^{2}+m_{2}^{2}$,
 $f=-m_{3}^{2}+i\epsilon$.
 
 Using the transformation cared out in reference [1]
 the integral (24 can be reduced to the following form:
$$
 R_{0}=\frac {i}{16\pi^{2}} \Biggl\{ \int \limits_{0}^{1}dy\frac {1}
 {(2\alpha b+c)y+d+e\alpha+2a+c\alpha}[\ell n(by^{2}+(c+e)y+d+a+f)-
$$
$$
 \ell n(by_{1}^{2}+(c+e)y_{1}+d+a+f)]-
 \int \limits_{0}^{1}dy\frac {1-\alpha}{(2\alpha b+c)y(1-\alpha)+d+e\alpha}
 [\ell n((a+b+c)y^{2}+(e+d)y+f)-
$$
$$
 \ell n((a+b+c)y^{2}_{2}+(e+d)y_{2}+f)]-
$$
$$
 \int \limits_{0}^{1}dy\frac {\alpha}{-(2\alpha b+c)y\alpha+d+e\alpha}
 [\ell n(ay^{2}+dy+f)-\ell n(ay_{3}^{2}+dy_{3}+f)]\Biggr\}, \eqno{(25)}
$$
 where
 $\alpha$ is the root of the quadratic polynomial $b\alpha^{2}+c\alpha+a$
 and $y_{1}$,~$y_{2}$,~$y_{3}$ are defined in the following way:
$$
 y_{0}=-\frac{d+e\alpha}{c+2b\alpha},~~~
 y_{1}=y_{0}+\alpha,~~y_{2}=\frac{y_{0}}{1-\alpha},~~~
 y_{3}=-\frac{y_{0}}{\alpha}. \eqno{(26)}
$$
 
>From (25) we see that all three integrals have the similar form. Let us
 consider the first of them:
$$
 F=\frac {i}{16\pi^{2}(2b\alpha+c)}\int \limits_{0}^{1}dy\frac{1}
 {y-y_{1}}\Biggl\{\ell n(by^{2}+(c+e)y+d+a+f)-
$$
$$
 \ell n(by_{1}^{2}+(c+e)y_{1}+d+a+f)\Biggr\}~. \eqno{(27)}
$$
 
 The function $F$ which is defined by formula (27) is expressed via the Spens
 functions [7-8]:
$$
 F=\frac {i}{16\pi^{2}(2b\alpha+c)}\Biggl\{ Sp\biggl(
 \frac {y_{1}}{y_{1}-y_{1}^{'}}\biggr)-
 Sp\biggl( \frac {y_{1}-1}{y_{1}-y_{1}^{'}}\biggr)+
 Sp\biggl( \frac {y_{1}}{y_{1}-y_{2}^{'}}\biggr)-
 Sp\biggl( \frac {y_{1}-1}{y_{1}-y_{2}^{'}}\biggr)\Biggr\}=
$$
$$
\frac {i}{16\pi^{2}(2b\alpha+c)}\sum_{i=1}^{2}\Biggl\{ Sp\biggl(
 \frac {y_{1}}{y_{1}-y_{1}^{'}}\biggr)-
 Sp\biggl( \frac {y_{1}-1}{y_{1}-y_{1}^{'}}\biggr)\Biggr\} \eqno{(28)}
$$
 Using the expressions (27) and (25) the three point function
 $R_{0}$ takes the final form:
$$
 R_{0}=\frac {i}{16\pi^{2}(2b\alpha+c)}\sum_{i=1}^{3} \sum_{\sigma=\pm}
 (-1)^{1+j}\Biggl\{ Sp\biggl(
 \frac {y_{i}}{y_{i}-y_{i}^{\sigma}}\biggr)-
 Sp\biggl( \frac {y_{i}-1}{y_{i}-y_{i}^{\sigma}}\biggr)\Biggr\}, \eqno{(29)}
$$
 where $y_{i}^{\pm}$ are the roots of the following quadratic
 polynomials respectively:
$$
 (p_{1}-p_{2})^{2}y^{2}+(m_{1}^{2}-m_{2}^{2}-(p_{1}-p_{2})^{2})y+m_{2}^{2}
 -i\epsilon,
$$
$$
 p_{1}^{2}y^{2}+(m_{1}^{2}-m_{3}^{2}-p_{1}^{2})y+m_{3}^{2}-i\epsilon,
$$
$$
 p_{2}^{2}y^{2}+(m_{2}^{2}-m_{3}^{2}-p_{2}^{2})y+m_{3}^{2}-i\epsilon.
$$
 
 It should be pointed out that with the using formulae
 (11)-(14),(20),(22),(29) all three point functions can be analyzed
 analytically us well as numerically.
 
 In the particular case, when one of the mass parameter significantly
 exit ales, it is more convenient to use the following expressions for
 the integrals:
$$
 R_{0}(p^{2}_{1},p^{2}_{2},(p_{1}-p_{2})^{2},
 m^{2}_{1},m^{2}_{1},m^{2}_{2})=-\frac {i}{(4\pi)^{2}}\frac {1}{m_{1}^{2}}
 \biggl\{ c_{1}(x)+\frac {p_{1}^{2}+p_{2}^{2}}{m_{1}^{2}}c_{2}(x)+
 \frac {(p_{1}-p_{2})^{2}}{m_{1}^{2}}c_{3}(x)
$$
$$
 R_{10}(p^{2}_{1},p^{2}_{2},(p_{1}-p_{2})^{2},
 m^{2}_{1},m^{2}_{1},m^{2}_{2})=-\frac {i}{(4\pi)^{2}}\frac {1}{m_{1}^{2}}
 \biggl\{ d_{1}(x)+\frac {p_{1}^{2}+2p_{2}^{2}}{m_{1}^{2}}d_{2}(x)+
 \frac {(p_{1}-p_{2})^{2}}{m_{1}^{2}}d_{3}(x)
$$
$$
 R_{01}(p^{2}_{1},p^{2}_{2},(p_{1}-p_{2})^{2},
 m^{2}_{1},m^{2}_{1},m^{2}_{2})=
 R_{10}(p^{2}_{2},p^{2}_{1},(p_{1}-p_{2})^{2},
 m^{2}_{1},m^{2}_{1},m^{2}_{2})
$$
$$
 \tilde R_{00}(p^{2}_{1},p^{2}_{2},(p_{1}-p_{2})^{2},
 m^{2}_{1},m^{2}_{1},m^{2}_{2})=-\frac {i}{(4\pi)^{2}}
 \biggl\{ \frac {1}{4}\ell n\frac {m_{1}^{2}}{\mu^{2}}+
 e_{1}(x)+\frac {p_{1}^{2}+p_{2}^{2}}{m_{1}^{2}}e_{2}(x)+
 \frac {(p_{1}-p_{2})^{2}}{m_{1}^{2}}e_{3}(x)
$$
$$
 R_{20}(p^{2}_{1},p^{2}_{2},(p_{1}-p_{2})^{2},
 m^{2}_{1},m^{2}_{1},m^{2}_{2})=-\frac {i}{(4\pi)^{2}}\frac {1}{m_{1}^{2}}
 \biggl\{ f_{1}(x)+\frac {p_{1}^{2}+3p_{2}^{2}}{m_{1}^{2}}f_{2}(x)+
 \frac {(p_{1}-p_{2})^{2}}{m_{1}^{2}}f_{3}(x)
$$
$$
 R_{11}(p^{2}_{1},p^{2}_{2},(p_{1}-p_{2})^{2},
 m^{2}_{1},m^{2}_{1},m^{2}_{2})=-\frac {i}{(4\pi)^{2}}\frac {1}{m_{1}^{2}}
 \biggl\{ \frac {1}{2}f_{1}(x)+
 \frac {p_{1}^{2}+p_{2}^{2}}{m_{1}^{2}}f_{2}(x)+
 \frac {2}{3}\frac {(p_{1}-p_{2})^{2}}{m_{1}^{2}}f_{3}(x)
$$
$$
 R_{02}(p^{2}_{1},p^{2}_{2},(p_{1}-p_{2})^{2},
 m^{2}_{1},m^{2}_{1},m^{2}_{2})=
 R_{20}(p^{2}_{2},p^{2}_{1},(p_{1}-p_{2})^{2},
 m^{2}_{1},m^{2}_{1},m^{2}_{2}), \eqno{(30)}
$$
 Where we have introduced the following notation:
$$
 c_{1}(x)=\frac {1-x+x\ell nx}{(1-x)^{2}}~,~~~~~
 c_{2}(x)=\frac {1+4x-5x^{2}+2x(2+x)\ell nx}{4(1-x)^{4}}~,
$$
$$
 c_{3}(x)=\frac {1-6x+3x^{2}+2x^{3}-6x^{2}\ell nx}
 {12(1-x)^{4}}~,~~
 d_{1}(x)=\frac {1-4x+3x^{2}-2x^{2}\ell nx}{4(1-x)^{3}}~,
$$
$$
 d_{2}(x)=\frac {1-9x-9x^{2}+17x^{3}-6x^{2}(3+x)\ell nx}{36(1-x)^{5}}~,~
 d_{3}(x)=\frac {1-6x+18x^{2}-10x^{3}-3x^{4}+12x^{3}\ell nx}{36(1-x)^{5}}~,
$$
$$
 e_{1}(x)=\frac {-1+4x-3x^{2}+2x^{2}\ell nx}{8(1-x)^{2}}~,~~
 e_{2}(x)=-\frac {1}{2}c_{3}(x),~~
 e_{3}(x)=\frac {-2+9x-18x^{2}+11x^{3}-6x^{3}\ell nx}{72(1-x)^{4}}~,
$$
$$
 f_{1}(x)=-4e_{3}(x)~,~~~
 f_{2}(x)=\frac {1-8x+36x^{2}+8x^{3}-37x^{4}+12x^{3}(4+x)\ell nx}
 {144(1-x)^{6}}~,
$$
$$
 f_{3}(x)=\frac {3-20x+60x^{2}-120x^{3}+65x^{4}+12x^{5}-60x^{4}\ell nx}
 {240(1-x)^{6}}~,~~~~x=\frac {m_{2}^{2}}{m_{1}^{2}} \eqno{(31)}
$$
 
 It should be pointed out that the expressions (17') and (31) at
 x=1 are considered as the limiting expressions for $x \rightarrow 1$.
\begin{center}
\section { SUMMARY. }
\end{center}
 
 In the given work the general expressions for the two and three
 point functions are obtained. The asymptotical behavior with respect to the
 values of the external and internal partical momenta is investigated.
 Different asymptotic representation for these functions (17), (18), (30) are
 given. It is convenient to use the different representations for these
 functions depending upon the comcret physical beyond investigated.
 Some particular cases it is possible to use integral representation
 (5)-(10), (22),(25) to written down answer in terms of one integral and them
 calculate this integral analyticaly or numericaly. In other casses it may
 be more convenoent to use the expressions (13)-(14), (22),(29).
 However,if the cinematic of the preces beyond investigated allows us to use
 the definite approximatins [9-10], then it is convenient from the practical
 point of weu to use the asymptotical expressions for the two and three
 point functions (17),(18),(30).
 
 The autors  expres their deep graduate to J. Gegelia and G. Japaridze
 for usefull discussions.


\newpage
\begin{thebibliography}{99}
\bibitem{1}
  G.t'Hooft and M.Veltman// Nucl. Phys. {\bf{B153}} (1979) 365.
\bibitem{2}
   G.Passarino and M.Veltman// Nucl. Phys. {\bf{B160}} (1979) 151.
\bibitem{3}
  Bernd A. Kniehl// Phys. Rep. {\bf{v.240}} (1994) 288.
\bibitem{4}
  D.Yu.Bardin, P.Ch.Christova and O.M.Fedorenko// Nucl. Phys.
{\bf{B197}}(1982) 1.
\bibitem{5}
 J.Gegelia, K.Turashvili//TMF 101 (1994) 225.
\bibitem{6}
 G.t'Hooft, M.Veltman// Nucl. Phys. {\bf{B44}} (1972) 189.
\bibitem{7}
 M.Abramovich, I.Stigan// Nauka, 1979.
\bibitem{8}
 L.Lewin. Dilogarithms and Associated functions (North-Holland,
 Amsterdam, 1981).
\bibitem{9}
 N.N.Bogolubov, D.V.Shirkov// Nauka, 1984.
\bibitem{10}
 A.I.Axiezer, V.B.Berestetskyi//Nauka, 1981.
\end{thebibliography}
\end{document}